\documentclass[floatfix,aps,onecolumn]{revtex4}
\usepackage[T1]{fontenc}
\usepackage{graphicx}

\newcommand{\beq}{\begin{equation}}
\newcommand{\eeq}{\end{equation}}
\newcommand{\bea}{\begin{eqnarray}}
\newcommand{\eea}{\end{eqnarray}}
\newcommand{\nn}{\nonumber}

\newcommand{\eps}{\epsilon}
\newcommand{\veps}{\varepsilon}
\newcommand{\al}{\alpha}

\newcommand{\D}{\Delta}

\newcommand{\be}{\beta}

\newcommand{\ra}{\rangle}
\newcommand{\la}{\langle}

\newcommand{\ga}{\gamma}
\newcommand{\om}{\omega}

\begin{document}

\title{Spectroscopic analysis of finite size effects around a Kondo quantum dot}
\author{Pascal Simon$^1$, and Denis Feinberg$^2$}
\affiliation{
$^1$ Laboratoire de Physique et Mod\'elisation des Milieux
     Condens\'es, Universit\'e Joseph Fourier and CNRS, BP 166, 38042 Grenoble, France\\
$^2$ Institut NEEL, CNRS and Universit\'e Joseph Fourier, BP 166, 38042 Grenoble, 
France}
%
%

\begin{abstract}
We consider a simple setup in which a small quantum dot is strongly connected to a 
finite size box. This box can be either a metallic box or a finite size quantum wire. 
The formation of the Kondo screening cloud in the box  strongly depends
on the ratio between the Kondo temperature and the box level spacing.
By weakly connecting two metallic reservoirs to the quantum dot, a detailed 
spectroscopic analysis can be performed.
Since the transport channels and the screening channels are almost decoupled,
such a setup allows an easier access to the measure of finite-size 
effects associated with the finite extension of the Kondo cloud.
\end{abstract}

\maketitle
\section{Introduction}
\label{intro}
The Kondo effect occurs as soon a magnetic impurity is coupled to a Fermi sea.
It is characterized by a narrow resonance of width $T_K^0$, the Kondo temperature,
pinned at the 
Fermi energy $E_F$ \cite{hewson}. This resonance is related to the many-body
singlet state which is formed between the impurity spin and the spin
of an electron belonging to a cloud of spin-correlated electrons.
This cloud of electrons has been termed as the so-called Kondo screening cloud.
The Kondo screening cloud length is therefore related to the spatial extension
of this multi-electronic, spin-correlated, wave function. The 
size of this screening cloud may be evaluated as $\xi_K^0\approx \hbar v_F/T_K^0$ where 
$v_F$ is the Fermi velocity. Nonetheless, the
Kondo screening cloud has never been detected experimentally and has
therefore remained a rather elusive prediction.

However, the remarkable recent achievements in nano-electronics
may offer new possibilities to finally observe this Kondo cloud.
Indeed, the Kondo effect appears as a rather robust and versatile phenomenon.
It has been observed by various groups  
in a single semi-conductor 
quantum dot \cite{dot,Cronenwett,simmel,Wiel}, in carbon nanotubes quantum dots \cite{cobden,bachtold,jarillo}, in molecular transistors \cite{molecular}
 to list a few. In this respect, the observation of the Kondo effect may be regarded as a test
of quantum coherence of the nanoscopic system under study. One of the main signatures
of the Kondo effect is a zero-bias anomaly and the conductance reaching the unitary
limit $2e^2/h$ at low enough temperature $T<T_K^0$. 
 
In a semi-conducting quantum dot, the typical Kondo temperature is of order
$1~K$ which leads to $\xi_K^0\approx 1$ micron in semiconducting heterostructures.
Finite size effects (FSE) related to the actual extent of this length scale have been 
predicted
recently in  different geometries: an impurity embedded in a finite size box \cite{thimm,monien},
a quantum dot embedded in a ring threaded by a magnetic flux
\cite{affleck01,Hu,sorensen04,simon05}, a quantum dot embedded 
between two open finite size wires (OFSW) (by open we mean 
connected to at least one external infinite lead) \cite{simon02,balseiro} and also around a double
quantum dot \cite{simon05b}.
In the ring geometry, it was shown that the persistent 
current induced by a magnetic flux is particularly sensitive to screening 
cloud effects and is drastically reduced when the circumference 
of the ring becomes smaller than $\xi_K^0$ \cite{affleck01}. In the wire geometry, 
a signature of the finite size extension  of the Kondo cloud was found in the temperature dependence of the 
 conductance through the whole system \cite{simon02,balseiro}. 
More specifically, in a one-dimensional 
 geometry where the finite size $l$ is associated to a level spacing $\D\sim \hbar v_F/l$, 
the Kondo cloud fully develops 
 if $\xi_K^0 \ll l$, a condition equivalent to $T_K^0 \gg \D$. On the contrary, 
FSE effects appear if 
 $\xi_K^0 > l$ or $T_K^0 < \D$. In a one dimensional geometry, one can equivalently use the ratio
$\xi_K^0/l$ or $\D/T_K^0$. In higher dimensions, we should rather use the latter ratio or
introduce another typical box length scale \cite{monien}.
In the aforementioned two-terminal geometry, the screening of the artificial spin impurity is done in the OFSWs which are also used to probe
transport properties through the whole system. This brings at least two main drawbacks: 
first, the analysis of FSE relies on the independent 
control of the two wire gate voltages and also a  rather symmetric geometry. This 
is rather difficult to achieve experimentally. 
In order to remedy to these drawbacks, we propose and study here a simpler setup in which 
the screening of the impurity occurs mainly in one larger quantum dot or metallic box\footnote{ Note that for a $2D$ or $3D$ metallic box, the level spacing $\Delta$ is not simply related to the Fermi velocity and the 
length scale of the box. In this case, we shall compare directly $T_K^0$ to $\Delta$.} or OFSW 
and the transport is analyzed by help of one or two weakly coupled leads. 
In practice, a lead weakly coupled to the dot by a tunnel junction allows a 
spectroscopic analysis of the dot local density of states (LDOS) in a way very similar to a STM tip.

The geometry we study is
depicted in Fig. \ref{Fig:device}. We note that this geometry has also been proposed by Oreg and
Goldhaber-Gordon \cite{oreg} to look for signatures of the two-channel Kondo fixed point or by Craig et al. \cite{craig} to analyze two quantum dots coupled to a common larger quantum dots and interacting {\it via} 
the RKKY interaction.
In the former case, the key ingredient is the Coulomb interaction of the box whereas in the latter, the box is largely open and used simply as a metallic reservoir mediating both the Kondo and RKKY interactions.
Here we are interested in the case where finite-size effects in the larger quantum dot or metallic
box do matter whereas in the aforementioned experiments the level spacing was among the smallest scales. Nevertheless, we emphasize that by increasing the box level spacing, the regime discussed in 
this paper 
should be accessible by these type of experiments.

The plan of the paper is as follows: in section 2, we present the model Hamiltonian
and derive how the FSE renormalizes
the Kondo temperature in our geometry.  In section 3, we perform a detailed spectroscopic analysis.
In section 4, we show how FSE affect the transport properties of the quantum dot. In section 5, the effect of a finite Coulomb energy in the box is discussed. Finally section 6 summarizes our results.

\label{sec:1}
\section{Model Hamiltonian and Kondo temperature}

\begin{figure}[b]
\centerline{\includegraphics[width=6.cm,,height=6.cm,clip=true]{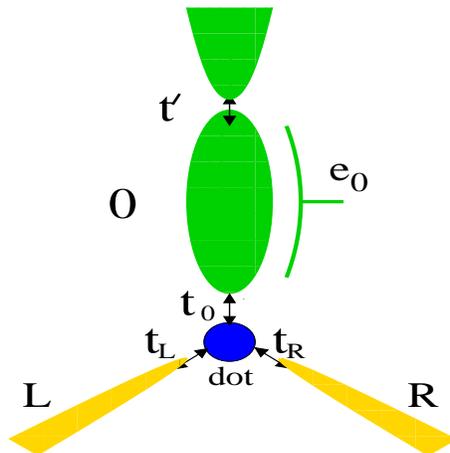}}
\caption{Schematic representation of the device we analyze in this paper. 
When the box Coulomb blockade energy is not neglected, 
we assume that the box potential can be controlled by a voltage gate $e_0$.}
\label{Fig:device}
\end{figure}

The geometry we analyze is depicted in Fig. \ref{Fig:device}. In this section we
assume that the large dot is connected to a third lead. From hereon, the Coulomb interaction
in the box is neglected except in section 5. 
The Coulomb interaction does not affect
the main results we discuss in this section.
In order to model the finite-size box connected to a normal reservoir, we choose for convenience
a finite-size wire characterized by its length  $l$ or equivalently by its level spacing $\Delta\sim \hbar v_F/l$. 
In fact, the precise shape of the finite-size box is not important for our purpose as soon as it is 
characterized by  a mean level spacing $\Delta$ separating peaks in the electronic density of states.
We assume that the small quantum dot is weakly coupled to one or two adjacent leads ($L$ and $R$). 
On the Hamiltonian level, we use the following tight-binding description, and for simplicity 
model the leads as one-dimensional wires (this is by no means restrictive): $H=H_L+H_R+H_0+H_{dot}+H_{tun}$ with
\bea\label{hamil}
H_{L}&=&-t\sum_{j=1,s}^{\infty} (c^\dagger_{j,s,L}c_{j+1,s,L}+h.c.)
-\mu_{L}  n_{j,s,L}  \\
H_0&=& -t\sum\limits_{j=1,s}^{\infty} (c^\dagger_{j,s,0}c_{j+1,s,0}+h.c.)-\mu_0 n_{j,s,0}\\
&&+(t-t')\sum_s (c^\dagger_{l,s,0}c_{l+1,s,0}+h.c.)\nn\\
H_{dot}&=&\sum_s \epsilon_{d} n_{d,s} +Un_{d\uparrow}n_{d\downarrow}\\
H_{tun}&=&\sum_s\sum_{\al=L,R,0}( t_\al c^\dagger_{ds} c_{1,s,\al}+h.c.).
\eea
$H_R$ is obtained from $H_L$ by changing $L\to R$.
Here $c_{j,s,\al}$ destroys an electron of spin $s$ at site $j$ in lead 
$\al=0,L,R$; $c_{d,s}$ destroys an electron with spin $s$ in the dot, $ n_{j,s,\al}= c^\dagger_{j,s,\al}c_{j,s,\al}$ and $n_{ds}= c^\dagger_{ds}c_{ds}$. The quantum dot is described by an Anderson impurity model, $\eps_{d},U$ are respectively the energy level and the Coulomb repulsion energy in the dot.
The tunneling amplitudes between the dot and the left lead, right lead and box are respectively 
denoted as $t_L,t_R,t_0$ (see Fig. 1). The tunneling amplitude amplitude between the box and the third lead is denoted
as $t'$ (see Fig. 1). Finally $t$ denotes the tight binding amplitude for conduction electrons 
implying that the electronic bandwidth $\Lambda_0=4t$. 
Since we want to use the left and right leads just as transport probes, we
assume in the rest of the paper that $t_L,t_R \ll t_0$.

We are particularly interested in the Kondo regime where $\la n_d\ra\sim 1$.
In this regime, we can map $H_{tun}+H_{dot}$ to a Kondo Hamiltonian by help of a Schrieffer-Wolff transformation:
\beq\label{hkondo}
H_K=H_{tun}+H_{dot}=\sum_{\alpha,\beta=L,R,0}
J_{\alpha\beta}c_{1,s,\alpha}^{\dagger}\frac{\vec{\sigma}_{ss'}}{2}\cdot\vec{S}c_{1,s',\beta},
\end{equation}
where $J_{\al\beta}=2t_\al t_\be(1/|\veps_d|+1/(\veps_d+U))$.
It is clear that $J_{00}\gg J_{0L},J_{0R}\gg J_{LL},J_{RR},J_{LR}$.
In Eq. (\ref{hkondo}), we have neglected direct potential scattering terms
which do  not renormalize and can be omitted in the low energy limit.

The Kondo temperature is a crossover scale separating the high temperature perturbative regime from the low temperature 
one where the impurity is screened.
There are many ways to define such scale. We choose the ``perturbative scale'' which is defined as the scale at which the 
second order corrections to the Kondo couplings become of the same order 
of the bare Kondo coupling. Note that all various definitions of Kondo scales differ by a constant multiplicative factor (see for example Ref. \cite{simon05} for a comparison between the perturbative Kondo scale with the one coming from the Slave Boson Mean Field Theory).

The renormalization group (RG) equations relate the Kondo couplings defined at scales $\Lambda_0$ and $\Lambda$. They simply
read:
\beq \label{RG0}
J_{\al\be}(\Lambda)\approx J_{\al\be}(\Lambda_0)
+{1\over 2}\sum_\ga J_{\al\ga}(\Lambda_0)J_{\ga\be}(\Lambda_0)
\left[\int\limits_\Lambda^{\Lambda_0}+\int\limits_{-\Lambda_0}^{-\Lambda}\right] {\rho_\ga(\om)\over |\om|}d\om\nn
\eeq
where $\rho_\gamma$ is the LDOS in lead $\ga$ seen by the quantum dot.
Since $J_{00}\gg J_{LL},J_{RR}$, the Kondo temperature essentially depends on the LDOS in the lead $0$.
Using the RG equations in Eq. (\ref{RG0}), the Kondo 
temperature  can be well approximated as follows:
\beq\label{deftk}
{ J_{00}\over 2}\left[\int\limits_{T_K}^{\Lambda_0}+\int\limits_{-\Lambda_0}^{-T_K}\right] {\rho_0(\om)\over |\om|}d\om=1
\eeq
When the lead $0$ becomes infinite (i.e. when $t'=t$), $\rho_0(\om)=\rho_0=const$ and we recover $T_K=T_K^0$
the usual Kondo temperature. 
It is worth noting that including the Coulomb interaction in the box does not affect much the Kondo temperature. 
The box
Coulomb energy $E_B$ slightly renormalizes $J_{00}$ in the Schrieffer-Wolff transformation 
since 
$E_{G}\ll U,|\eps_d|$ \cite{oreg}.

The LDOS $\rho_0$ can be easily computed for a finite one-dimensional wire.
In general, for a finite size open structure, $\rho_0$
corresponds in the limit of a weak coupling to a continuum
to a sum of resonance peaks. The positions of these peaks is
to a good approximation related to the eigenvalues $\om_n$ of the isolated finite size structure,
while their width $\ga_n$ is proportionnal to $t'^2|\psi_n|^2$ where the $\psi_n$ are the eigenvectors of the isolated structure.
The LDOS $\rho_0$
is very well approximated by a sum of Lorentzian functions\cite{simon02} in the limit $t'\ll t$ :
\beq\label{lorentzienne}
\pi \rho_0(\om)\approx \sum_n |\psi_n|^2 \frac{\gamma_n}{(\om-\om_n)^2+\gamma_n^2}.
\eeq
For a $1D$ finite size wire of length $l$, $\psi_n=2\sin(k_n)/(l+1)$ with $k_n\approx \pi n/(l+1)$.
This approximation is quite convenient in order to estimate the Kondo temperature $T_K$ through (\ref{deftk}). 
 
When the level spacing $\Delta_n \sim \hbar v_F/l$ is much smaller
than the Kondo temperature $T_K^0$, no finite-size effects are expected. Indeed, the integral in (\ref{deftk})
 averages out over many peaks and the genuine Kondo temperature is $T_K\sim T_K^0$.
On the other hand, when $T_K^0 \sim\Delta_n $, the Kondo temperature starts to depend on the fine structure 
of the LDOS $\rho_0$ and a careful calculation of the integral in
(\ref{deftk}) is required. Two cases may be distinguished: either $\rho_0$ is tuned such that 
a resonance  $\om_n$ sits at the Fermi energy $E_F=0$ (labeled by the index $R$) or in a non resonant situation  
(labeled by the index $NR$). 
In the former case, we can estimate
\bea\label{def:tkr}
T_K^R&=&\frac{\gamma_n \Delta_n}{\sqrt{(\Delta_n^2+\gamma_n^2)\exp\left(\frac{2}
 {J_{00}(\Delta_n )\rho_0^R(0)}\right)-\Delta_n^2}}\nn \\
&\approx&
\gamma_n\exp\left(-\frac{1}
 {J_{00}(\Delta_n)\rho_0^R(0)}\right)\approx \ga_n\left(\frac{T_K^0}{\D_n}\right)^{\frac{\pi\ga_n}{\D_n}}.
\eea

In the latter case, we obtain,
\bea\label{def:tknr}
T_K^{NR}&=&\Delta_n \exp\left(-\frac{\pi \Delta_n^2}{4 J_{00}(\Delta_n)|\psi_n|^2\gamma_n} \right)\nn\\
&\approx& \Delta_n \exp\left(-\frac{1}{J_{00}(\Delta_n)\rho_0^{NR}(0)}\right)
\approx  \Delta_n \left(\frac{T_K^0}{\D_n}\right)^{\frac{\pi\D_n}{8\ga_n}}, 
\eea
where $\pi\rho_0^{NR}(0)\approx 4\gamma_n|\psi_n|^2/\Delta_n^2$.
These two scales are very different when $t'^2\ll t^2$. By controlling $\rho_0$, we can control the 
Kondo temperature (at least when $T_K\leq \D$).
The main feature of such geometry is that the screening of the artificial spin impurity 
is essentially 
performed in the open finite-size wire corresponding to lead $0$. Now let us study what are the consequences of FSE  on transport 
when one or two leads are weakly coupled to the dot. This is the purpose of the next section.

\section{Spectroscopy of a Kondo quantum dot coupled to an open box}\label{sec:spectro}

In this section, we consider a  standard three-terminal geometry as depicted in Fig. \ref{Fig:device}.
Since the leads are weakly connected to the dot, they allow a direct access to the dot density
of states in presence of FSE in the box.

We  have used the Slave Boson Mean Field
Theory (SBMFT) \cite{hewson}  in order to calculate the dot local density of states (LDOS).
This approximation describes qualitatively well the behavior of 
the Kondo impurity
at low temperature $T\leq T_K$ when the impurity is screened and especially capture well
the exponential dependence of the Kondo temperature.
Furthermore, this method has been proved to be efficient to capture finite size effects 
in Refs \cite{simon02,simon05}.

\begin{figure}
\includegraphics[width=6.5cm,height=6.cm,angle=-90,clip=true]{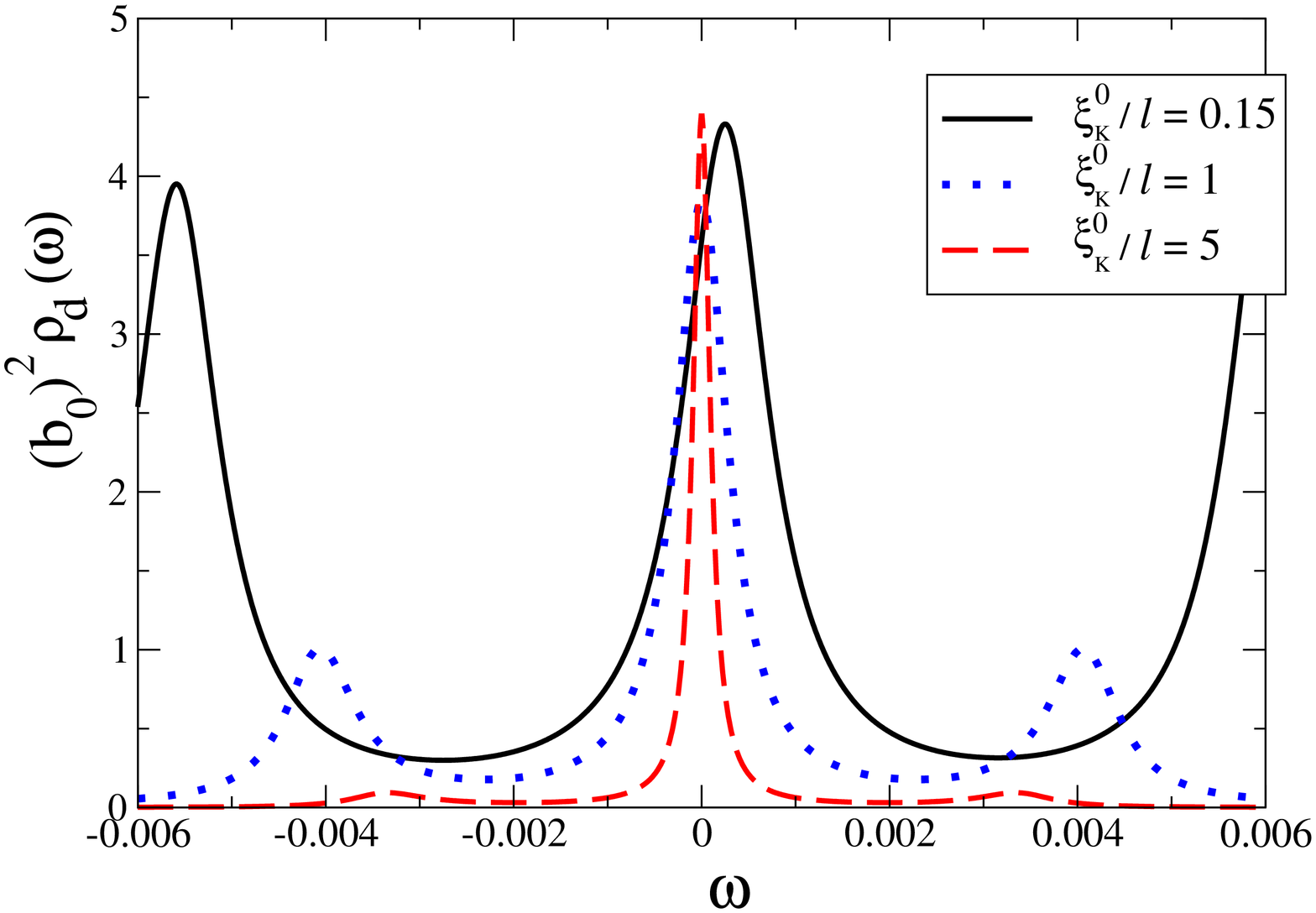}
\includegraphics[width=6.5cm,height=6.cm,angle=-90,clip=true]{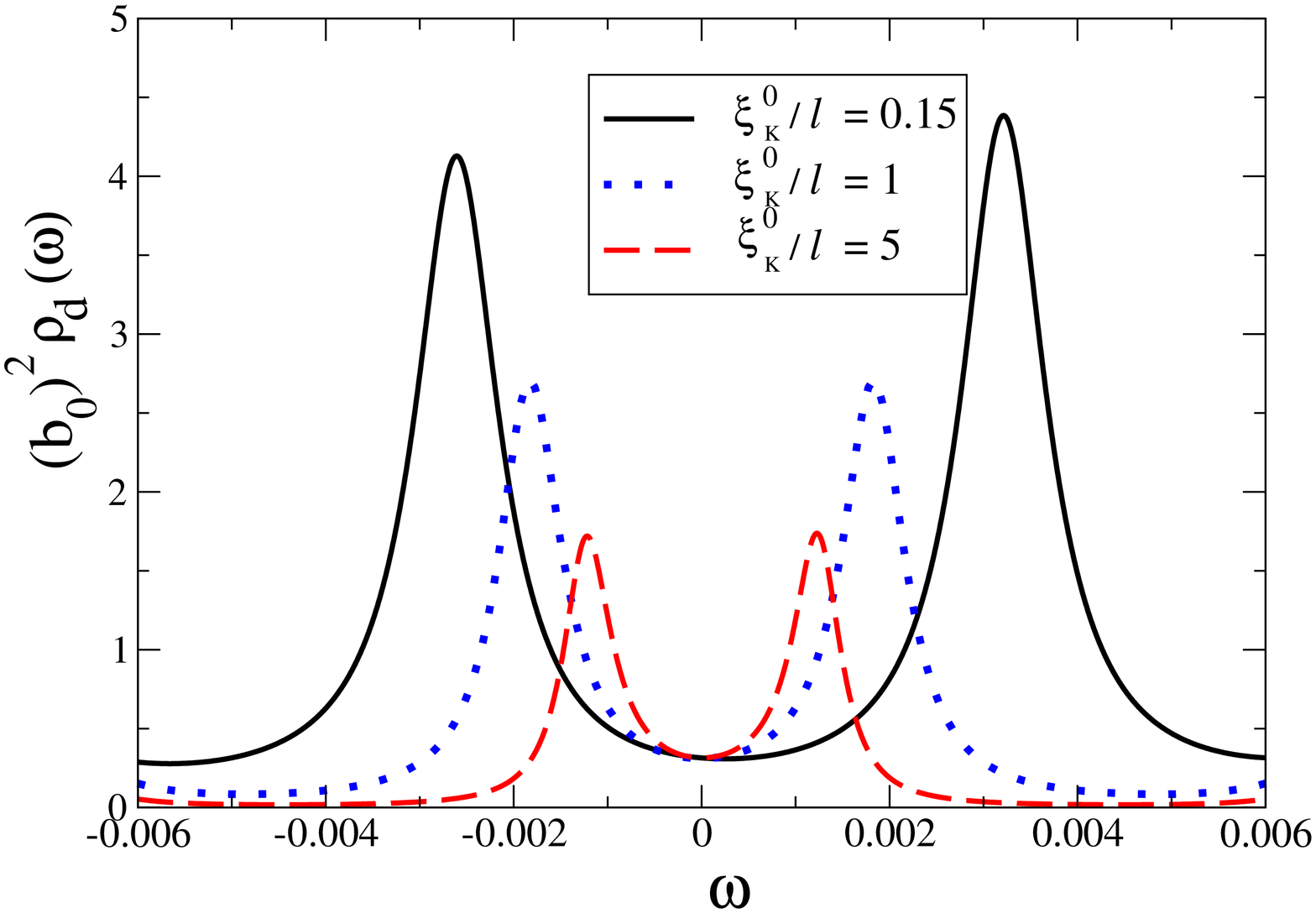}
\caption{Dot density of states $\rho_d(\omega)$ for  both the non resonant case (upper panel)
and the resonant case (lower panel). We took $\Delta\sim 0.006$ and  plot $\rho_d$ for $\xi_K^0/l=\Delta/T_K^0\sim 0.15 $, $\xi_K^0/l=\Delta/T_K^0\sim 1$ and $\xi_K^0/l=\Delta/T_K^0\sim 5$. 
Note that $\rho_d$ has been scaled by $b_0^2$, the slave boson parameter for an easy comparison between both cases. }
\label{Fig:rhod}
\end{figure}

Let us now compute the dot density of states $\rho_d(\omega)$.
We assuming $\Gamma_{L/R}\ll \Gamma_0(\om)$ 
such that for low bias the dot Green functions weakly depend on
the chemical potential in the left and right leads.
Under such conditions, the differential conductance reads as follows:
\beq
e\frac{dI}{d\mu_L}\approx\frac{2e^2}{h}4\Gamma_L\int\limits_{-\infty}^\infty\left(\frac{-df(\om)}{d\om}\right)
\pi\rho_d(\om+\mu_L) d\om.
\eeq
Varying $\mu_L$ allows a direct experimental access to $\rho_d(\mu_L)$ at $T\ll T_K$.
Note that a similar approximation is used for STM theory with magnetic adatoms \cite{schiller00}.
We have plotted $\rho_d(\omega)$ in Fig. \ref{Fig:rhod} for both the non-resonant case
and the on-resonance case for three different values of
$\xi_K^0/l$. We took the following parameters in units of $t=1$: $t_0=0.5$, 
$t_L=t_R=0.1$ (therefore $t_L^2<<t_0^2$), $t'=0.5$, and $l\sim 1000a$ ($a$ the lattice constant) or
equivalently $\Delta\sim 0.006$. 
The Kondo energy scale $\xi_K^0$ can be varied using the dot energy level $\eps_d$ which is controlled 
by the dot gate voltage.  One has to distinguish between two cases: $T_K^0 \gg \D$
and $ T_K^0 \ll \D$.

When $T_K^0 \gg \D$, no finite-size effect is to be expected. 
Nevertheless two non trivial features should be be noticed:
i) The various peaks appearing in $\rho_d$ are included in an envelope of 
width $O(T_K^0)\gg \Delta$
(which has a broader range than the figure \ref{Fig:rhod} actually covers for $\xi_K^0/l=\Delta/T_K^0\sim 0.15$). In this respect, the Kondo resonance plays the role of an energy filtering device
which filters the box high energy states that are not in the range of width $O(T_K^0)$
around $E_F=0$.

ii) The LDOS $\rho_d(\omega)$
mimics the density of state in the lead $0$ but is shifted  such that an off-resonance peak
in the lead  $0$ corresponds to a dot resonance peak and vice versa.
This can be simply understood from a non-interacting picture valid at $T=0$. 
The non-interacting dot Green function 
reads 
\beq
G_{dd}(\omega)\approx\frac{1}{\omega -\eps_d -\delta \eps(\omega)+i\Gamma_0(\omega)},\eeq
where $\delta\eps$ is the real part of the dot self-energy and $\Gamma_0$ its imaginary part.
The minima's of $\Gamma_0$ thus correspond to the maxima of $-Im(G_{dd})$. 
We also note that the resonance peak
is slightly shifted from $\om=0$ in this limit. Therefore the 2-terminal conductance does not reach its unitary limit (i.e. its maximum non interacting value). 
This is due to the fact that we took $\eps_d=-0.68$ and we are not deep in the Kondo regime.
Particle-hole symmetry is not completely restored in the low energy limit.
 
On the other hand, when $\xi_K^0\gg l$ ($T_K^0\ll \D$), $\rho_d$ changes drastically. 
The fine structure in the box density of states no longer shows up in $\rho_d$. 
This is expected since only the energy states which are within a range of order $O(T_K)\ll \Delta$
appears in the dot LDOS.
In the off-resonance case, only the narrow Kondo 
peak of width $T_K^{NR}\ll T_K^0$ mainly subsists for $\xi_K^0/l=5$ (upper panel of Fig. \ref{Fig:rhod}).
 We can also show that the position of the small peaks at $\omega\sim \pm\Delta/2$ for $\xi_K^0\gg l$ 
are related to the resonance peaks in lead $0$. We also note that the narrow peak is this time almost at $\om=E_F=0$
i.e. particle-hole symmetry is restored. In order to reach large value of $\xi_K^0$, we took small values
of $\eps_d$ such that we are deep in the Kondo regime where $n_D\sim 1$.

The most surprising result occurs for the resonant case
where the Kondo peak is split for $\xi_K^0\gg l$. Usually a splitting of the Kondo LDOS
is associated with the destruction of the Kondo effect as this would be the case for a 
Kondo quantum dot under a magnetic field or interacting with another quantum dot via the RKKY interaction \cite{craig,rkky,vavilov}. Here the splitting cannot at all be attributed to the destruction of the Kondo effect but instead to a subtle non interacting destructive interference phenomenon.
Let us use some renormalization group argument. When we integrate out high energy electronic 
degrees of 
freedom from the bandwidth  down to an energy scale of order $\D>T_K^0$, we start building the Kondo resonance at the Fermi energy $E_F$.  If we continue integrating out electronic degrees of freedom
from $\D$ down to $\ga$, this reasoning would tell us that we end up with a Kondo resonance pinned at $E_F$. However, this does not take into account that the box has also a (non interacting) resonance pinned at $E_F$ and these two resonances
are coupled via the tunnel amplitude $t_0$. By analogy to a molecular system
with two degenerate orbital states, where 
a tunneling amplitude leaves the degeneracy and 
creates  bonding and anti-bonding states, a strong $t_0$ (as is our case here) may lead to a splitting of the Kondo resonance.
Therefore this splitting is more related to a destructive interference between the two resonances -an interacting one in the dot and a non-interacting one in the box. 
The study of the splitting of the Kondo resonance as a function of $t_0$ has been 
extensively studied recently in a slightly different geometry by Dias da Silva et al. \cite{sandler}. 
One can also quantify this splitting within the SBMFT method.
By approximating 
$\Gamma_0(\omega)\approx t_0^2 |\psi_n|^2\frac{\gamma_n}{\omega^2+\gamma_n^2}$
 at a resonance $n$, 
one can well understand analytically
the structure of $\rho_d$. When $T_K^R\gg \ga_n$, 
one can  show that the peak splitting is of order 
$\sim 2\sqrt{\gamma_n T_K^R}$ and the peaks width is of order $\gamma_n$.

\section{Analysis of transport properties}
At $T=0$, it is straightforward to show, using for example the scattering formalism,\cite{ng} 
that the conductance matrix $G_{\al,\be}^U$ is simply given
by

\beq\label{giju}
G_{\al,\be}^U=\frac{2e^2}{h} {4\Gamma_\al\Gamma_\be\over (\Gamma_L+\Gamma_0+\Gamma_R)^2}
\eeq
where $\Gamma_\al=\pi t_\al^2\rho_\al(0)$, and $\al=l,0,r$.
Since the SBMFT aims at replacing the initial Anderson Hamiltonian by a non-interacting one,
one may easily access the conductance by directly applying the Landauer formula
or equivalently by using
\beq\label{mw1}
G_{\al\be}=\frac{2e^2}{h}\int d\omega \left(\frac{-\partial f}{\partial \omega}\right) 
 {4\Gamma_\al(\omega)\Gamma_\be(\omega)\over (\sum_\al \Gamma_\al(\omega))}
Im(-G_{dd}^r)(\omega).
\eeq
Using the SBMFT, one can extensively study the conductance as function of temperature for various case $T_K^0\gg \D$, $T_K^0\ll D$. This has been reported in details in \cite{julien}.
Actually, it turns out that finite size effects, which are related to the finite size extension of the Kondo cloud, are clearly visible at some  intermediate temperature $T_K^0> T>T_K^R,T_K^{NR}$. 
In this temperature range, significant deviations from 
the non-interacting limit are obtained. Let us assume that the box can be gated (see Fig. 1).
We took the same parameters as in Fig. 2 except that $l=200a$.
The finite size effects are much more spectacular when
one look at the conductance, at fixed temperature, as a function of the box gate voltage $e_0$. 
We have therefore fixed the temperature at $T=2.10^{-3}$ and plotted the conductance $G_{L0}$
as a function of $e_0$ 
for different
values of $\xi_K$, controlled here by the parameter $\eps_d$. The other parameters are unchanged.
The upper (plain style) curve corresponds to $\xi_K=50a<l$. We observe large oscillations of the conductance corresponding to $e_0$ being on a resonance or off a resonance. At large $\xi_K\sim 1000a\gg l$,
the conductance (dashed-dashed-dotted style) has a completely different shape. The minima's and the maxima's 
of the conductance at $\xi_K\sim 50a$  become  now respectively maxima's and minima's. Furthermore the conductance at these minima's is very small close to $0$. This regime corresponds to the high temperature for the non resonant case. 
The intermediate values of $\xi_K$ show how the conductance crosses over in between
these two extreme cases. This dramatic change of the conductance in the regime in which
$\xi_K\gg l$ is a direct consequence of interactions effects and should be directly  observable
in experiments. 
Similar results can be obviously obtained by analyzing the 
conductance $G_{LR}$ despite the fact that 
the amplitude of $G_{LR}$ will be in general smaller than $G_{L0}$ by a factor $\sim \sqrt{\Gamma_L\Gamma_R}/\Gamma_0$. In fact, the conductance $G_{LR}$ can be made significantly larger by adjusting the chemical potential $\mu_0$ such that the current in lead $0$ verifies $\langle I_0 \rangle=0$.

\begin{figure}
\includegraphics[width=7.5cm,height=8.5cm,angle=-90,clip=true]{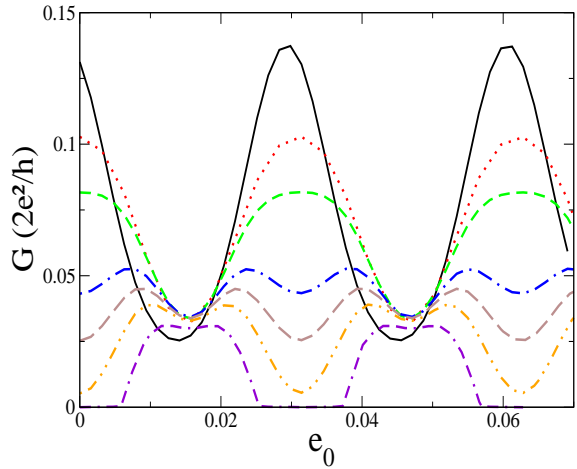}
\caption{Conductance (in units of $2e^2/h$) as a function of $\veps_W$ for different values of $\xi_K$. From top to bottom
$\xi_K\sim 50a$ (plain style), $\xi_k\sim 150a$ (dotted style), $\xi_K\sim 300a$ (short dashed style), $\xi_K\sim 500a$ (dot-long dashed style), $\xi_K\sim 600a$ (long dashed style), $\xi_K\sim 750a$ (dot-dot-dashed style) and $\xi_K\sim 1000a$ (dot-dashed-dashed style)} 
\label{Fig:Gew}
\end{figure}
We have essentially used the SBMFT to analyze the spectroscopic and transport properties.
Nevertheless, one can also use analytical calculations in two limiting cases.
When $T\gg T_K$, one can safely rely on renormalized perturbative calculations while at $T\ll T_K$
the Nozi\`eres Fermi liquid approach \cite{nozieres} can be applied. Using the latter theory, one can
for example show that the conductance in the on resonance case is a non monotonous function of temperature. We refer the reader to Ref. \cite{julien} where these analytical calculations are detailed and confirm the present analysis.

\section{Finite box Coulomb energy}
In this section, we discuss whether  a finite box Coulomb energy  modifies
or not the results presented in this work. As we already mentioned in section 2, the Kondo
coupling $J_{00}$, is almost not affected by the box Coulomb energy $E_B$ (since $E_B\ll U$)
and therefore the Kondo temperature remains almost unchanged.
As shown in \cite{oreg,borda04,florens},  a small energy scale $E_B$ changes the 
renormalization group equation
in Eq. (\ref{RG0}). The off-diagonal couplings $J_{0L}(\Lambda),J_{0R}(\Lambda)$ tend to $0$ 
for $\Lambda\ll E_B$.
At energy $\Lambda\ll E_B$ 
the problem therefore reduces to an anisotropic $2-$channel Kondo problem. 
The strongly coupled channel is the box $0$, the weakly coupled one is the even combination
of the conduction electron in the left/right leads. At very low energy, the fixed 
point of the anisotropic $2-$channel Kondo model is a Fermi liquid. It 
is characterized by the strongly coupled lead (here the box)
screening the impurity whereas the weakly coupled one completely decoupled from the impurity.
The dot density of states depicted in Fig. \ref{Fig:rhod} should remain therefore almost unaffected.
The problem is to read the dot LDOS with the weakly coupled leads
since they decouple at $T=0$. Nevertheless, for a typical experiment done at low
temperature $T$, such a decoupling is not complete and the dot LDOS should be still 
accessible using the weakly coupled leads but with a very small amplitude.

We up to now analyze the situation in which a box or a finite size wire is also
used as a third terminal {\it i.e} is coupled to a continuum. In some situations, 
like the theoretical one presented 
in Ref. \cite{oreg}, no terminal lead is attached to the box
and the geometry is a genuine 2-terminal one. In order to analyze this system, 
we have to take into account 
both a finite level spacing and a finite box Coulomb energy.
One can make progress if we assume $T_{K0}\ll\Delta\ll E_B\ll U\ll D_0$ where $D_0$ is the bandwidth.
One proceeds with a RG treatment in three steps:  $E_B<<\Lambda<<U$, $\Delta<<\Lambda<<E_B$ et $T_{K0}<<\Lambda<<\Delta$. 
When  $E_B<<\Lambda<<U$, the RG equations derived in Eq. (\ref{RG0}) remain valid and are the usual ones. When  $\Delta << \Lambda << E_B$, the couplings $J_{L0}$ and $J_{R0}$ stop renormalizing because of the Coulomb energy $E_B$. The other couplings keep on growing. 
When $T_{K0}<<\Lambda<<\Delta$, two situations are to be considered. Let us start with the case where
no resonance sits at the Fermi energy in the box. The coupling $J_{00}$ no longer renormalizes since there is no state available. On the other hand, the couplings $J_{\al,\be}$ with $\al,\be=L,R$
keep on renormalizing and may eventually reach a strong coupling regime. Therefore, the screening
of the impurity ultimately occurs in the $L,R$ leads though we may have started with $J_{00}\gg J_{LL},J_{RR}$. This situation is quite different than the one we explored previously.
However, when a resonance sits at the Fermi energy of the box, we are left with a double dot problem
with one electron in each dots. They form a singlet and we expect a split LDOS in the dot. 
If we assume $E_B\ll \D$, the problem gets more complicated since we already 
reach a strong coupling regime at the scale $E_B$.
A step toward this direction was recently achieved in Ref. \cite{kaul06} in a slightly modified geometry.

\section{Conclusions}

In this paper, we have studied a geometry in which a small quantum dot in
the Kondo regime
is strongly coupled to a large open quantum dot or open finite size wire
 and weakly coupled to other normal leads which are simply used as transport probes.
The artificial impurity is mainly screened is the large quantum dot. Such a geometry 
thus allows to probe the dot spectroscopic properties without perturbing it. 
We have shown using the SBMFT how finite size effects show up in 
the dot density of states and in the the conductance
matrix. We also analyzed how these results are modified in presence of 
Coulomb interactions in the box.
We hope the predictions presented here are robust enough to be 
checked experimentally.

{\bf Acknowledgements}
This research was partly supported by the contract PNANO `QuSpins' of the French Agence Nationale de la Recherche.


\begin{thebibliography}{}
\bibitem{note0} Note that for a $2D$ or $3D$ metallic box, the level spacing $\Delta$ is not simply related to the Fermi velocity and the 
length scale of the box. In this case, we shall compare directly $T_K^0$ to $\Delta$.
\end{thebibliography}
\end{document}